\documentclass[a4paper]{llncs}

\usepackage{amssymb}
\usepackage{graphicx}

\usepackage{url}
\urldef{\mailsa}\path| jmanuel@bacchuss.com.ar|

\usepackage[utf8]{inputenc}  
\usepackage[spanish]{babel}

\begin{document}

\mainmatter  

\title{Ciberseguridad Inteligente}


%
%
\author{Juan Manuel R. Mosso}
\authorrunning{Lecture Notes in Computer Science: Authors' Instructions}

\institute{Bacchuss, División Ciberdefensa\\
Sarmiento 537 PB B., 2000 Rosario, Santa Fe, Argentina\\
\mailsa\\
\url{http://www.bacchuss.com.ar}}

%
%

\toctitle{Lecture Notes in Computer Science}
\tocauthor{Authors' Instructions}
\maketitle

\begin{abstract}
Los modelos de negocios en la economía moderna poseen una fuerte dependencia del ciberespacio. Esta situación plantea escenarios de riesgo de seguridad debido a la cada vez mayor cantidad de vulnerabilidades y al incremento en la frecuencia y en la sofisticación de los ciberataques, especialmente con el advenimiento de amenazas avanzadas con características de persistencia conocidas como APT. El presente trabajo presenta un modelo de Ciberseguridad Inteligente (CSI) destinado a detectar, denegar, interrumpir, degradar, engañar y destruir las capacidades del adversario en el ciberespacio. Esto se logra por medio del desarrollo conceptual y técnico de una Capacidad de Ciberinteligencia (CCI) cuyo objetivo es el de interferir las capacidades de planificación y operación del adversario, penetrando sus bucles de decisión con la velocidad necesaria, a fin de lograr su desplazamiento a una postura reactiva. Finalmente, a diferencia de los modelos de ciberseguridad  planteados clásicamente, el concepto de CSI sugiere que la ventaja en un escenario de conflicto puede ser obtenida por el defensor y no siempre por el atacante. Como sustento teórico se presentan el "Modelo de Sistema Ofensivo de Referencia" (MSOR), el cual es utilizado para pensar a la ciberseguridad de manera integral, a todo nivel, desde una perspectiva coordinada y sincronizada con el resto de las áreas con responsabilidades directas o indirectas en temas de ciberseguridad. Finalmente se esboza una justificación del modelo planteado desde la perspectiva moderna de los sistemas de control. El presente trabajo resulta de especial interés para organizaciones de los sectores público y privado vinculadas al sistema de infraestructura crítica nacional. \newline

{\bfseries Palabras clave:} Ciberseguridad, ciberdefensa, ciberinteligencia, inteligencia, defensa nacional, infraestructura crítica.

\end{abstract}

\section{Introducción}

Las organizaciones públicas y privadas modernas poseen una fuerte dependencia del ciberespacio como sustento de sus modelos de negocio. Esta situación plantea escenarios de riesgo de seguridad debido a la cada vez mayor cantidad de vulnerabilidades y al incremento en la frecuencia y en la sofisticación de los ciberataques dirigidos contra componentes de infraestructura crítica que no reconoce fronteras entre lo civil y lo militar, lo público y lo privado, lo local y lo transnacional. 

Esta realidad, marcada por el amplio uso del ciberespacio, por su complejidad y dinamismo, responde a la revolución que posibilitó el inicio de la transformación del comercio internacional a partir de la década de 1980 como respuesta a limitaciones tecnológicas fundamentales existentes al momento. En la actualidad, todo modelo de negocio surge de la interconexión a varios niveles de diversos componentes en lo que puede pensarse como una forma de Comercio Centrado en Redes (CCR) ó en la economía de Internet. La esencia de la transformación se sustentó inicialmente en el concepto de que la información puede ser transformada en una ventaja competitiva para lograr la superioridad y el liderazgo comercial. El concepto de CCR ha influenciado profundamente la manera de hacer negocios alrededor del mundo. \newline

Después de dos décadas de evolución, la economía de Internet parece estar en jaque debido a la temática de seguridad. Más aún, la ciberseguridad se ha convertido en la principal amenaza al comercio en el ciberespacio por lo que las naciones del mundo se enfrentan a la necesidad de dar respuesta a la problemática, no solo desde lo tecnológico, sino desde lo político y lo legal. Esta nueva realidad queda evidenciada por diferentes eventos mundiales como lo muestra la Organización para la Cooperación y el Desarrollo (OECD) \cite{b1}, la presentación de Enero de 2014 de Toomas Hendrik Ilves, presidente de Estonia, en la Conferencia de Seguridad de Munich \cite{b2}, y los artículos de Lucas Kello \cite{b3} y de Erik Gartzke \cite{b4} en el MIT Press Journal de Seguridad Internacional de 2013. En este contexto, y con el fin de garantizar la continuidad de las operaciones de los diferentes sistemas que dan soporte al comercio de Internet se propone el desarrollo del concepto Ciberseguridad Inteligente (CSI). Este se apoya en el desarrollo de un Capacidad de Ciberinteligencia (CCI) cuyo rol es el de complementar y salvar las limitaciones de los modelos tradicionales de ciberseguridad como la seguridad operativa y la respuesta a incidentes de seguridad en redes de cómputo conocidas como CERT/CSIRT. A tal fin se presenta el "Modelo de Sistema Ofensivo de Referencia" (MSOR), recurso conceptual desarrollado con el objetivo de contar con una herramienta base para el análisis, el diseño, el desarrollo, la operación y el mantenimiento de la CCI, aplicable a cualquier tipo de escenario y a todo nivel, estratégico, operacional y táctico. \newline

Es importante considerar que a lo largo del trabajo se utilizará cierto tipo de lenguaje específico del dominio militar como \textit{campañas, operaciones, conflicto, técnicas, tácticas y procedimientos (TTP)} el cuál, por la naturaleza del trabajo, aplica de igual manera al dominio civil.

\section{Interés público}

El presente trabajo constituye un aporte teórico destinado a sustentar el desarrollo de capacidades de ciberseguridad para organizaciones con requerimientos fuertes de seguridad de la información. Resulta de especial interés para su aplicación en organizaciones vinculadas directamente con la temática de infraestructuras crítica nacional  \cite{b5} en base a lo establecido por el “Programa Nacional de Infraestructuras Críticas de Información y Ciberseguridad” (ICIC) creado mediante la Resolución JGM Nº 580/2011 \cite{b6}.

\section{Internet como blanco}

En esta sección se analizan conceptos fundamentales relacionados con el ciberespacio como blanco potencial para diferentes tipos de amenazas. Debido a su carácter estratégico y a la transversalidad de la información en todos los dominios de la sociedad moderna, la ciberseguridad se vuelve un factor clave. Un aspecto importante a considerar es la dependencia de cada vez más procesos y funciones de misión crítica vinculadas al desempeño de servicios y funciones elementales para la vida de las sociedades modernas. Esta dependencia trasciende fronteras geográficas y geopolíticas, civiles y militares, públicas y privadas. El ciberespacio está distribuido mundialmente y se encuentra integrado a las Infraestructuras Críticas (IC) de las diversas naciones del mundo, vinculado de manera directa con su gestión y operación. El grado de desarrollo y dependencia sobre este hacen que resulte un activo clave para el desarrollo de aspectos tan fundamentales como el gobierno, el comercio, y la seguridad entre otros. \newline

\emph{Ciberespacio:} Entorno complejo que resulta de la interacción de las personas, el software y los servicios a través de Internet por medio de dispositivos tecnológicos y redes conectadas a esta, que no posee entidad física \cite{b7}. \newline

El ciberespacio, por su importancia, es actualmente sujeto de operaciones de tipo ofensivas cuya motivación varía según el tipo de amenaza, a saber:

\begin{enumerate}
\item {\bfseries Estados Nación:} Este tipo de amenaza es la más peligrosa debido a la capacidad de acceso a recursos en períodos prolongados de tiempo. Las naciones pueden hacer uso de operaciones en el ciberespacio (OC) con el objetivo de atacar recursos de IC o desarrollar espionaje. En este contexto deben ser considerados los adversarios tradicionales, los que podrán ejecutar las operaciones de manera directa o a través de servicios contratados a terceros.
\item {\bfseries Grupos Organizados Transnacionales: } Conformadas por organizaciones que no se encuentran vinculadas a ninguna nación, con la capacidad de actuar a nivel global. Este tipo de actores utiliza el ciberespacio para desarrollar sus actividades ilícitas como robo, fraude, e incluso para desarrollar operaciones de desestabilización de gobiernos y ataques terroristas. En este sentido, el ciberespacio sirve de plataforma para la planificación y sincronización de las operaciones irregulares, y para la comunicación entre sus componentes. Este tipo de amenaza es utilizada por naciones de mundo u organizaciones privadas para desarrollar ataques o espionaje mediante OC.
\item {\bfseries 	Pequeños grupos o individuos:} Este tipo de amenaza en general se manifiesta en términos de interrupción de servicios o de accesos no autorizados a recursos clasificados. Pueden ser utilizados por estados nación ú organizaciones criminales para externalizar  operaciones y así deslindar responsabilidad. Existen tantas motivaciones como individuos, algunas de las cuales pueden sostener posiciones maliciosas. Un ejemplos de esto lo constituye el ciberactivismo.  
\item {\bfseries 	Amenaza interna:} Conformada por individuos con privilegios lícitos de acceso a los recursos de información de una organización. Al igual que con actores individuales, estos poseen infinitas motivaciones. 
\end{enumerate}

\subsection{El rol del sector privado}

Muchos de los recursos en base a los cuales se sostiene la ciberseguridad y la defensa nacional se sostienen en infraestructuras que son propiedad o que se encuentran operadas por organizaciones del sector privado. Este hecho plantea al menos dos efectos que deben ser atendidos. El primero de ellos enfocado en la necesidad de crear conciencia en los proveedores de servicios sobre aspectos de seguridad con el fin de hacerlos parte en estrategias colectivas de gestión de riesgos. El segundo se debe a la necesidad de poder contar con los recursos privados vinculados a las operaciones de infraestructura crítica en operaciones conjuntas con el objetivo de responder a incidentes de seguridad. 

\subsection{Tipos de operaciones}

El logro de objetivos en el contexto de seguridad en el ciberespacio contempla la ejecución de operaciones ofensivas, defensivas y de inteligencia.

\begin{enumerate}
\item {\bfseries Operaciones Ofensivas:} Tienen como objetivo proyectar poder en y a través del ciberespacio. Se basan en acciones destinadas a evitar que el adversario haga uso de los recursos (ej.: degradación, interrupción o destrucción), o a obtener el control o a la modificación de los mismos.
\item {\bfseries Operaciones Defensivas: } Tienen como objetivo  defender el ciberespacio propio y el de los aliados, preservando la funcionalidad de todas las capacidades y servicios necesarios. Existen dos tipos:
\begin{itemize}
\item Ciberdefensa Pasiva: Es la defensa del ciberespacio basada principalmente en la aplicación de controles de seguridad dentro de los perímetros de las redes y sistemas propios.
\item Ciberdefensa Activa: Tomando la definición de Dewar \cite{b8}, y extendiendo los conceptos al nivel estratégico, puede establecerse que la ciberdefensa activa constituye un enfoque por medio del cual se puede lograr la seguridad del ciberespacio por medio del despliegue de medidas destinadas a detectar, analizar, identificar y mitigar las amenazas hacia y desde los sistemas de comunicaciones y redes, en escalas de tiempo que van desde lo instantáneo (tiempo real) hasta años, combinado con la capacidad y los recursos para tomar medidas proactivas u ofensivas contra las amenazas en sus propias redes.
\end{itemize}
\item {\bfseries 	Operaciones de Inteligencia (CI):} Acciones destinadas a generar el conocimiento necesario para soportar operaciones militares en curso o como soporte de misiones a futuro. Además, sobre la base de contrainteligencia, desinforma o engaña al adversario sobre diferentes componentes del MSOR.
\end{enumerate}

\subsection{Comando y Control (C2)}

El Comando y Control es la función que conlleva el ejercicio de la autoridad y la dirección de las fuerzas por parte de las autoridades vinculadas a las operaciones en los diferentes ámbitos de TIC en vías del cumplimiento de una misión. En este sentido, el acceso a información exacta y en tiempo favorece la percepción de los responsables y aumenta la calidad del proceso de toma de decisiones.

\subsection{Consideraciones sobre la selección de blancos}

La selección de blancos en el ciberespacio es un proceso destinado a identificar y priorizar blancos potenciales en vías del cumplimiento de una misión. La selección y la determinación de las acciones a ejecutar por parte de la amenazas contra estos dependerá de la motivación, de los recursos y las capacidades operativas. El objetivo final del proceso de selección de blancos consiste en lograr la integración y sincronización de la transferencia de efectos de poder. La naturaleza del ciberespacio plantea desafíos fundamentales en lo que hace a la selección de blancos para el desarrollo de operaciones de tipo defensivas debido a las implicancias políticas internacionales y a la generación de efectos colaterales y en cadena que pueden generarse.

\section{Inteligencia en el ciberespacio}

El logro de los objetivos de seguridad en el ciberespacio requiere de la realización de suposiciones en relación a las condiciones operativas vigentes en el mismo. En este contexto, la inteligencia en el ciberespacio (CIBERINTEL) aporta los recursos necesarios para confirmar o corregir los supuestos y así afectar significativamente los niveles de riesgo asociados al escenario operacional.

\subsection{Consideraciones conceptuales}

La CIBERINTEL constituye un recurso de naturaleza compleja que, a pesar de no contar aún con una definición ni con un marco conceptual unificado, resulta imprescindible como herramienta en el contexto de ciberseguridad tanto en el plano estratégico como en el operativo y el táctico. En este sentido, la inteligencia debe derivarse de los planos físico, lógico y cognitivo por medio de las técnicas que resulten convenientes. Como tal, y para el caso de análisis en el ciberespacio, debe poder ser sujeta a esquemas conceptuales y metodológicos que permitan abordar la problemática involucrando a todos los niveles de decisión, y pueda aportarse una vista común de la situación. \newline

La CIBERINTEL representa una disciplina analítica la cual se nutre de datos de diversos tipos de fuentes para generar información y conocimiento en base a evidencias. Esta debe trascender lo puramente tecnológico y absorber factores como el  humano y el geopolítico, respondiendo a fuentes de información tanto internas como externas.

\subsection{Consideraciones técnicas}

A los fines del presente trabajo, bastará con presentar un modelo básico para el desarrollo de inteligencia basado en las definiciones de indicadores y precursores. 
\newline

\emph{Indicador:} Pieza de información asociada a la ocurrencia de un evento determinado con impacto en el modelo de negocios. \newline \newline
\emph{Precursor:} Pieza de información vinculada a la potencial ocurrencia de un evento con impacto.
\newline

La capacidad de desarrollar inteligencia se basa entonces en la identificación y obtención de un conjunto de indicadores y precursores por medio de procesos de identificación, agregación, análisis, fusión y presentación de información con un fuerte componentes de colaboración. Los indicadores y precursores deben poder ser sujetos a: 

\begin{itemize}
\item Trazabilidad: Proceso por medio del cuál tanto los indicadores como los precursores deben poder ser medidos en su evolución. 
\item Validación: Proceso por medio del cual, para un determinado instante, se logra establecer la validez del valor asociado a un indicador o precursor en función del conocimiento adquirido hasta ese preciso instante.
\end{itemize}

La trazabilidad y la validación de indicadores y precursores constituyen recursos de valor fundamental para la CIBERINTEL.
\newline

\emph{Capacidad de Ciberinteligencia (CCI):} Recurso organizacional de carácter estratégico basado en un conjunto de herramientas, técnicas, tácticas y procedimientos que, al ser explotados por personal calificado, permiten transformar la información obtenida de los diversos indicadores y precursores en conocimiento. 
\newline

La CCI debe desarrollar los recursos necesarios para poder hacer frente al desafío de transformar información en conocimiento, tal como lo indica la siguiente tabla, en el contexto de las leyes y tratados nacionales e internacionales:

\begin{center}
\begin{tabular}{| p{6cm} | p{6cm} |}
\hline
\emph{Información}	&	\emph{Inteligencia}	\\ \hline
Cruda, alimentación sin filtros	&	Procesada y ordenada	\\ \hline
No evaluada al ser distribuida	&	Evaluada e interpretada por analistas de inteligencia entrenados	\\ \hline
Agregada desde cualquier fuente	&	Agregada desde fuentes confiables y verificada por medio de correlación	\\ \hline
Puede ser verdadera, falsa, engañosa, incompleta, relevante o irrelevante	&	Exacta, en tiempo, completa (en medida de lo posible), y evaluada para determinar su relevancia	\\ \hline
No accionable	&	Accionable	\\ \hline
\end{tabular}
\end{center}

\section{Modelo de Sistema Ofensivo de Referencia}

En el presente capítulo se presenta el "Modelo de Sistema Ofensivo de Referencia" (MSOR). El MSOR constituye un modelo teórico de procesos que destaca los aspectos funcionales y los flujos de información involucrados en escenarios de conflicto en sus tres niveles, estratégico, operacional y táctico. Como se podrá ver en el desarrollo del mismo, se hace uso de terminología que permite formalizar conceptos la cual deriva de conceptos utilizados en el dominio militar, aunque con aplicación directa a los ámbitos público y privado. Los procesos que conforman el MSOR son: \emph{1) Proceso de Surgimiento del Conflicto (PSC), 2) Proceso de Planificación Estratégica (PPE), 3) Proceso de Ejecución de Operaciones (POC), y 4) Proceso de Evaluación de Resultados (PER)}. A continuación, en la Figura 1 se muestra la estructura lógica del MSOR y se desarrollará cada uno de los procesos en el orden previsto por el sistema.

\begin{figure}
\centering
\includegraphics[scale=1]{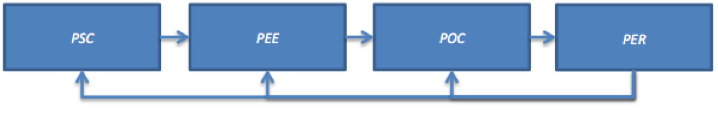}
\caption{Modelo de Sistema Ofensivo de Referencia" (MSOR)}
\end{figure}

\subsection{Proceso de Surgimiento del Conflicto (PSC)}

En la Figura 2 puede verse una representación del proceso PSC.

\begin{figure}
\centering
\includegraphics[width=\linewidth]{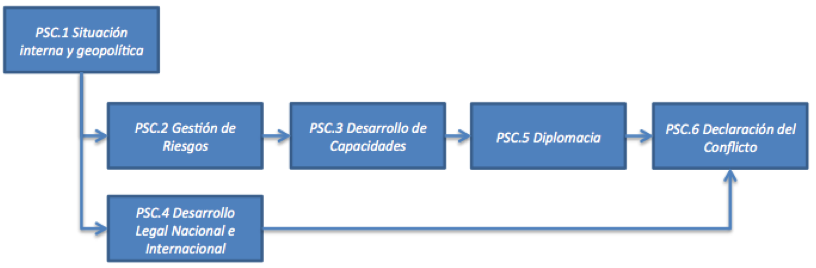}
\caption{Proceso de Surgimiento del Conflicto (PSC)}
\end{figure}

\textbf{PSC.1 Situación interna y geopolítica} 

Conceptualmente, el MSOR toma un estado inicial de equilibrio definido por la inexistencia de conflicto. En este contexto, tanto los servicios de soporte como las capacidades existentes se encuentran preparadas para actuar y en permanente actualización. \newline

\textbf{PSC.2 Gestión de Riesgos}	

La apertura del ciberespacio hace que deba realizarse una reconceptualización de la seguridad en sus diferentes niveles, nacional, regional, y global. La gestión de riesgos se consolida como una herramienta estratégica destinada a reducir la incertidumbre y la impredecibilidad definida por las relaciones y las tensiones internas y externas. La gestión de riesgos es un proceso que posibilita el tratamiento y la disminución de las vulnerabilidades frente a las  amenazas existentes. \newline

\textbf{PSC.3 Desarrollo de Servicios}

El desarrollo de servicios involucra la creación sistemas destinados a consolidar las políticas y los mecanismos de seguridad del ciberespacio con el fin de garantizar la continuidad del negocio. \newline

\textbf{PSC.4 Desarrollo Legal Nacional e Internacional}

Toda organización pública o privada, especialmente aquellas que conforman la infraestructura crítica nacional, deben desarrollar sus actividades en el marco jurídico existente, tanto a nivel nacional como internacional. \newline

\textbf{PSC.5 Diplomacia}

Desde el mismo momento en que se reconoce que gran parte del desarrollo social y económico del mundo moderno depende fuertemente del ciberespacio, la diplomacia comienza a jugar un rol fundamental en la disputa de intereses en este dominio. Las relaciones entre organizaciones de los sectores público y privado dependerán de las prioridades de cada una de ellas, en definitiva de sus intereses particulares. \newline

\textbf{PSC.6 Declaración del Conflicto}

En el MSOR, el surgimiento del conflicto de manera formal o informal constituye el evento que saca al sistema de la estabilidad. Este se da ante alguna de las siguientes hipótesis:

\begin{itemize}
\item Cuando los esfuerzos diplomáticos son agotados y la crisis entre las organizaciones no puede ser solucionada por vía de negociaciones pacíficas, ó
\item	Ante ataques irregulares.
\end{itemize}

La declaración del conflicto da inicio al segundo componente del MSOR, el proceso de PPE, en el que los niveles máximos de mando de las organizaciones inician la planificación de las aciones.

\subsection{Proceso de Planificación Estratégica (PPE)}

El PPE establece la forma general en la que se planifican, preparan, ejecutan y evalúan los resultados tanto a nivel estratégico como operacional de las actividades de aseguramiento del ciberespacio. Debido a las características de transversalidad del  ciberespacio en los que hace a infraestructura crítica, tanto pública como privada, el análisis de los resultados y de los efectos potenciales de toda acción debe ser determinado a priori, de manera minuciosa, apoyados en el conocimiento necesario que surja de la inteligencia aplicada al contexto definido. \newline

\emph{Planificación:} Proceso analítico que aporta los detalles sobre los recursos, las funciones y la sincronización entre las diferentes capacidades necesarios para cumplir con los objetivos de negocio a nivel organización con carácter local o global. \newline

El PPE, tomado del proceso de planificación conjunta de las FF:AA de los EE:UU por su simplicidad y alcance \cite{b9}, permite la transformación de objetivos estratégicos en actividades que se manifiestan en términos de proyectos y tareas sustentados en las capacidades operacionales establecidas. En la Figura 3 se muestra una representación del PPE:

\begin{figure}
\centering
\includegraphics[width=\linewidth]{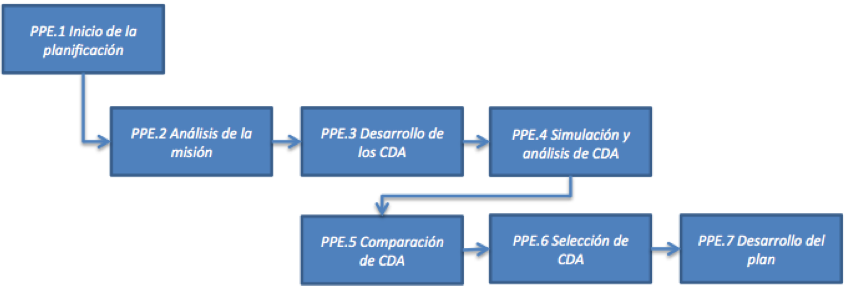}
\caption{Proceso de Planificación Estratégica (PSC)}
\end{figure}

A continuación se presentan las consideraciones asociadas a cada fase del PPE. \newline

\textbf{PPE.1 Inicio de la planificación}

La integración de las capacidades tecnológicas y del resto de las áreas organizativas vinculadas de manera directa e indirecta debe ser iniciada en la fase inicial de planificación. Entre las principales acciones a desarrollar se destacan el análisis de la misión, la revisión de documentos estratégicos, el análisis de los objetivos y la determinación del alcance inicial de las operaciones, y la identificación de blancos potenciales derivados del proceso de gestión de riesgos, entre otras. \newline

\textbf{PPE.2 Análisis de la misión}
 
El propósito de esta fase consiste en lograr una definición clara de la misión y una evaluación de los requerimientos a nivel organización. Desde la perspectiva operativa, como el ciberespacio cruza a muchos procesos de negocio, resultará fundamental mantener el foco en los objetivos planteados originalmente en la planificación. Algunas de las acciones claves de la fase de análisis son la identificación de fuerzas aliadas y de los adversarios, la identificación de factores críticos para la misión, la identificación de los factores relevantes de los dominios físico, de información y cognitivo en el ciberespacio, y el establecimiento del criterio de éxito de la misión, entre otros. \newline

\textbf{PPE.3 Desarrollo de los cursos de acción}

En el desarrollo de los cursos de acción (CDA) se explotan los productos derivados de la fase  de análisis. Entre las funciones clave de esta fase se destacan la identificación de los efectos deseados definidos por la misión al más alto nivel organizativo y a nivel operacional, los efectos indeseados que pudiesen impactar de manera negativa en los objetivos, y el análisis e identificación de los riesgos y controles complementarios, entre otros. \newline

\textbf{PPE.4 Simulación y análisis de CDA }

En base al tiempo disponible se deberá simular cada propuesta de CDA en consideración. Estas serán entonces contrastadas contra los CDA potenciales reconocidos del adversario mediante la inteligencia adquirida. Algunas de las acciones destacadas son el análisis de cada CDA contra los requerimiento funcionales a nivel táctico, la redefinición o ajuste de los requerimientos de inteligencia, el establecimiento de la información relevante para las operaciones, y la evaluación del riesgo correspondiente, entre otros. \newline

\textbf{PPE.5 Comparación de CDA }

Esta fase requiere del análisis y la evaluación de los diferentes CDA por parte de todo el personal involucrado, de las diferentes áreas intervinientes. Se determinan y evalúan las fortalezas y debilidades de cada CDA desde las diferentes perspectivas presentes. Las acciones más significativas en esta fase son el análisis y comparación de cada CDA contra la misión y las tareas, y su priorización. \newline

\textbf{PPE.6 Selección de CDA }

En esta fase, el personal relacionado con las OC plantea una recomendación a los niveles gerenciales superiores, quizá a nivel de dirección, sobre la manera en la que el CDA seleccionado contribuye al logro de la misión. Es fundamental que la recomendación sea realizada de manera clara y precisa para que pueda ser comprendida por todos los involucrados en la toma de decisiones. \newline

\textbf{PPE.7 Desarrollo del plan}

Seleccionado y aprobado un CDA, el personal vinculado a las OC desarrolla el plan. Este deberá considerar las cuestiones de integración con todas las áreas relacionadas a nivel organización para garantizar la efectividad del mismo. Las principales acciones de esta fase son la identificación de las deficiencias de las capacidades involucradas en el CDA y la recomendación de las soluciones, la indicación a los gerentes y directores sobre posibles temas que afecten o puedan afectar a las OC, entre otros. \newline

\subsection{Proceso de Operaciones en el Ciberespacio (POC) }

La etapa de ejecución de las OC se sustenta en las capacidades desarrolladas y establecidas a nivel organización y en la coordinación global de recursos a nivel organización en un teatro de operaciones integrado y sincronizado en base al plan establecido. En este contexto, toda acción llevada adelante en el ciberespacio deberá estar sometida a la autoridad relevante la cuál dirigirá las tareas para lograr sinergia entre los diferentes participantes. \newline

\emph{Operación:} Una secuencia de acciones a nivel táctico con el objetivo de cumplir un objetivo común. Dominio de acción que se desarrolla en el nivel táctico. \newline

\emph{Campaña:} Un conjunto de operaciones destinadas al cumplimiento de objetivos operacionales y estratégicos en un dado tiempo y espacio. Dominio de acción que se desarrolla en el nivel operacional. \newline

Para modelar el POC se propone un modelo basado en una adaptación del Cyber Killchain de Hutchins, Cloppert y Amin \cite{b10} y de la doctrina "F2T2EA" \cite{b11} utilizada por las FF:AA de los  EE:UU para la selección de objetivos. El POC puede ser pensado como una cadena de eventos centrado en una postura ofensiva, analogía que será de especial valor al momento de desarrollar estrategias de ciberseguridad. Así, el POC constituye un modelo elemental por medio del cual resulta posible abstraer los detalles de comportamiento asociados a amenazas, con énfasis en el nivel táctico. En la Figura 4 se presenta la estructura del proceso POC: \newline

\begin{figure}
\centering
\includegraphics[width=\linewidth]{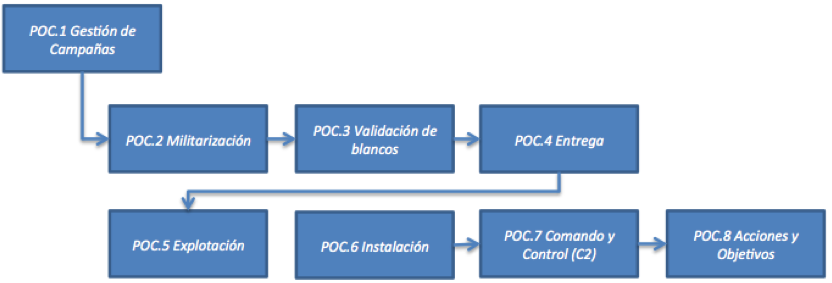}
\caption{Proceso de Operaciones en el Ciberespacio (POC)}
\end{figure}

\textbf{POC.1 Análisis de Campañas}

El análisis de campañas constituye un recurso analítico cuyo propósito consiste en determinar la manera en la que opera un adversario en el largo plazo. El análisis de campaña permite obtener información sobre los patrones ofensivos del adversario, es decir sobre sus TTP, herramientas, blancos y formas de proceder para lograr sus objetivos estratégicos. Esto implica la identificación de factores comunes en el análisis de indicadores y precursores sobre los cuales se desarrollará el conocimiento necesario para el diseño y selección de CDA defensivos adecuados. Ofrece respuestas al \textit{¿Cómo?} y \textit{¿Para qué?} por sobre el \textit{¿Qué?}. \newline

\textbf{POC.2 Militarización}
 
Este eslabón de la cadena complementa a las fases de desarrollo de capacidades del MSOR, y responde a demandas generadas por la dinámica del ciberespacio. En general son desarrollos de escala menor y a horizontes de tiempo considerablemente más cortos. Apunta a completar las capacidades de manera dinámica en base a los requerimientos específicos del ambiente particular. \newline
 
\textbf{POC.3 Validación de blancos} 

En base a los blancos seleccionados en las fases de planificación, este eslabón del proceso consiste en la generación de la inteligencia necesaria para su validación o para la modificación de la lista. Además se prevé el desarrollo de la inteligencia necesaria a fin de lograr el efecto deseado sobre estos. \newline

\textbf{POC.4 Entrega}

Este eslabón tiene por objetivo la transmisión del artefacto malicioso al ambiente objetivo. Los adjuntos de correo electrónico, los sitios Web, y los medios de almacenamiento removibles de tipo USB conforman los ejemplos de vectores de entrega más comunes. \newline

\textbf{POC.5 Explotación}

Una vez transmitido y entregado el artefacto, el eslabón de explotación dispara o inicia el código malicioso. Normalmente, la explotación consiste en la utilización de vulnerabilidades a nivel de sistema operativo o aplicaciones, de características o funcionalidades asociados a las TIC, y/ó mediante técnicas como el abuso de la confianza de usuarios por medio de ataques de ingeniería social. \newline

\textbf{POC.6 Instalación}

En general, este eslabón es el que permite establecer y consolidar la presencia en los ambientes de información del adversario. Consiste en la instalación de troyanos o puertas traseras, de tal manera de lograr acceso con características de persistencia en el ambiente objetivo. \newline

\textbf{POC.7 Comando y Control (C2)}

Una vez consumada la ejecución, la entidad comprometida envía una señal a un controlador por medio de Internet con el fin de establecer un canal de tipo C2. Una vez establecido el canal, el adversario posee acceso al ambiente objetivo para desarrollar el plan previsto. \newline

\textbf{POC.8 Acciones y Objetivos }

En este eslabón del POC, y luego de haber cumplido con éxito las primeras siete fases, el adversario posee el acceso necesario en el ambiente objetivo como para desarrollar el plan ofensivo y lograr sus objetivos. Los ataques pueden estar dirigidos a la obtención de resultados cibercinéticos directos o indirectos sobre componentes de infraestructura, de infraestructura crítica, o a la generación de plataformas de relevo en base a las cuales operar sobre otros objetivos locales o por medio de Internet. \newline

\textbf{Sobre la ciberdefensa proactiva}
 
Desde una perspectiva de defensa, del modelo POC surge claramente un quiebre en sus eslabones medios. La segunda mitad de la cadena, a partir del eslabón de explotación, contempla el conjunto de acciones vinculadas directamente a la obtención de los resultados por parte del adversario. Es sobre esta segunda mitad que se aplica el paradigma tradicional de ciberseguridad sustentado principalmente en los modelos tradicionales de gestión de incidentes de seguridad (CSIRT). \newline

Este paradigma de seguridad posee una limitación fundamental, su reactividad. Esta característica obedece al hecho de que la gestión de incidentes, la respuesta, presupone que el adversario ha logrado penetrar los sistemas de alguna manera, o que está en vías de hacerlo. En la Figura 5 se muestra el modelo POC y los paradigmas de seguridad  posibles planteados en base a lo expuesto.
 
\begin{figure}
\centering
\includegraphics[width=\linewidth]{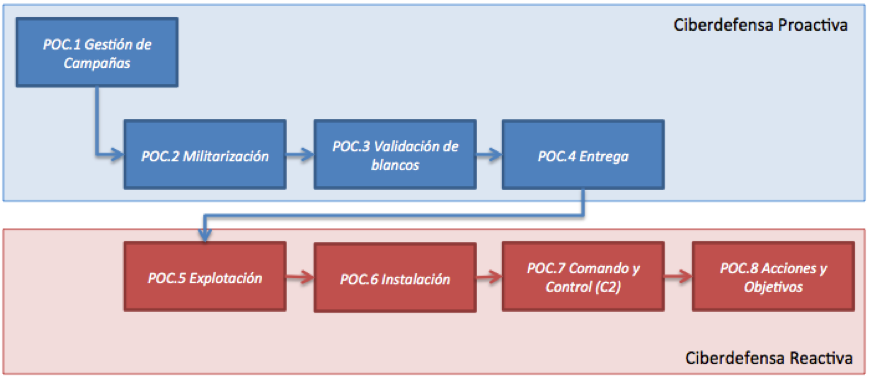}
\caption{POC y paradigmas de ciberseguridad}
\end{figure}

\subsection{Proceso de Evaluación de Resultados (PER)}

La evaluación de los resultados de las OC debe ser implementada como una herramienta a todos los niveles de conflicto en el ciberespacio, tanto en las fases de planificación como de ejecución. El proceso de evaluación se sustenta en la siguiente serie de conceptos: 1) Evaluación de los efectos de las operaciones y las campañas, 2) Determinación del grado de avance sobre los objetivos, 3) Mejoras para el desarrollo de futuras  acciones, 4) Producción de la realimentación necesaria para la toma de decisiones, y 5) Aportación de datos para la determinación del retorno de la inversión. En la Figura 6 se presenta la estructura del PER.

\begin{figure}
\centering
\includegraphics[scale=1]{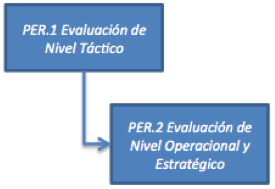}
\caption{Proceso de Evaluación de Resultados (PER)}
\end{figure} 

Las evaluaciones pueden ser implementadas a nivel táctico para determinar la efectividad de las operaciones, y a nivel operacional para establecer la efectividad de las estrategias implementadas y así poder generar recomendaciones a los diferentes niveles de mando. \newline

\textbf{PER.1 Evaluación de Nivel Táctico }

Consiste en la determinación de la efectividad de las operaciones tácticas en base a la recolección y el análisis de un conjunto de indicadores tácticos objetivos, de tipo cualitativo y cuantitativo, previamente definidos en las fases de planificación. El resultado de este tipo de evaluación indica a los mandos medios y altos de mando  la necesidad o no de desarrollar más acciones. La determinación del éxito de la misión y del efecto generado se logra por medio del análisis y la generación de inteligencia utilizando medios tales como SIGINT, GEOINT, OSINT, HUMINT y CIBERINTEL derivada. \newline

\textbf{PER.2 Evaluación de Nivel Operacional y Estratégico}

Brinda una herramienta destinada a soportar el juicio analítico de la estrategia implementada, es decir de los fines, las formas y los medios empleados para el logro de los objetivos. Permite determinar el progreso en el esfuerzo para lograr los objetivos operacionales y estratégicos planteados, y en base a este se genera información de realimentación destinada a ajustar la estrategia y los posibles CDA futuros.

\section{Ciberseguridad Inteligente}

Tanto los modelos MSOR y POC, como las consideraciones específicas de inteligencia y su aplicación al ciberespacio planteados en los capítulos anteriores son utilizados en el presente capítulo para presentar un marco teórico de ciberseguridad centrado en inteligencia, Ciberseguridad Inteligente (CSI). La CSI se sustenta en el diseño, la implementación, la operación y el mantenimiento de una capacidad de ciberinteligencia  como complemento de las capacidades tradicionales de protección del ciberespacio. La CSI logra su objetivo por medio de la creación de inteligencia producto de los Contextos de Información (CXI) derivados de cada una de las fases del MSOR. \newline

La base conceptual de la propuesta de CSI es motivada por la idea de desplazar al adversario hacia una postura reactiva en sus flujos de decisión en los niveles estratégicos, operacional y táctico del conflicto, y así obligarlo a ajustar de manera permanente sus objetivos, recursos y operaciones. Este cambio permanente en la dinámica de decisión derivará en mayores tiempos y costos vinculados a los diferentes procesos involucrados en el logro de su misión. Se cree que el modelo que sustenta la CCI plantea una alternativa viable a la máxima en seguridad: \textit{ "...el que ataca siempre posee la ventaja"}. A continuación se presenta un ejemplo sobre el cual se desarrolla el concepto que da entidad a la CSI y justifica el desarrollo de la CCI como recurso necesario para la ciberseguridad. 

\subsection{Nivel táctico del ciberconflicto}

\emph{Nivel táctico:} "Nivel en el que se planifican y ejecutan las operaciones destinadas a conducir al logro de los objetivos por parte de las unidades tácticas, es decir del personal técnico especializado de campo. Se trata de la organización y maniobra de los elementos de acción en relación a sus pares, y para con el adversario”\cite{b12}. \newline

Constituye el nivel en el que se desarrollan las acciones y en el que se genera el impacto sobre la infraestructura del adversario. Aquí es donde la acción se desarrolla, donde se enfrentan defensores y atacantes empleando sus arsenales, donde se explotan las vulnerabilidades. El nivel táctico, desde la perspectiva de ciberseguridad, presupone que el adversario ya ha logrado establecer presencia o que se encuentra muy próximo a lograrlo. Es el nivel del ciberconflicto que se vincula con la segunda parte del POC y que conceptualmente se ejecuta a la velocidad de la red, es decir en el que las acciones se consuman en el orden de milisegundos. El nivel táctico del ciberconflicto se relaciona con los eslabones: a) POC.3 Entrega, b) POC.4 Explotación, c) POC.5 Instalación, d) POC.6 Comando y Control (C2), y e) POC.7 Acciones sobre los objetivos; del modelo MSOR. La Figura 7 muestra un posible CDA basado en acciones de defensa. \newline

\begin{figure}
\centering
\includegraphics[scale=1]{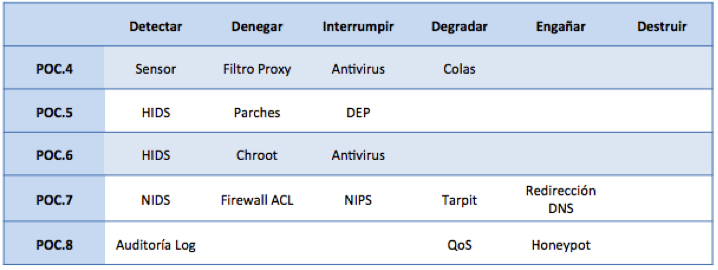}
\caption{CDA a nivel táctico (I)}
\end{figure} 

El desarrollo del los CDA apropiados para cada situación se desprenderá de la inteligencia que surja al explotar los contextos de información de nivel táctico.

\subsection{Nivel operacional del ciberconflicto}
\emph{Nivel operacional:} “Nivel en el que se planifican, conducen y mantienen las campañas u operaciones mayores con el fin de lograr los objetivos estratégicos”. \newline

La decisión de un adversario sobre la aplicación de una acción ofensiva, la planificación y la logística necesarias, y las capacidades y los recursos necesarios para llevarla adelante requieren de una serie de actividades que distan mucho de ser automáticas y de ejecutarse a velocidades cercanas a la de la red. Estas requieren de una dimensión de tiempo diferente, mayor, la cuál puede ser de días, semanas e incluso años para el caso de las APT. La posibilidad de combinar datos del pasado en base a la evidencia registrada a nivel de red con datos relativos a la situación local y global, a los conflictos regionales existentes, a las potenciales decisiones o acciones que tomará el adversario, a sus capacidades y objetivos finales, permitirá obtener la inteligencia necesaria para operar y obtener ventaja competitiva. Esta visión es la que justifica el desarrollo de inteligencia en los niveles operacional y estratégico. La Figura 8 muestra los eslabones correspondientes al nivel operacional.

\begin{figure}
\centering
\includegraphics[scale=1]{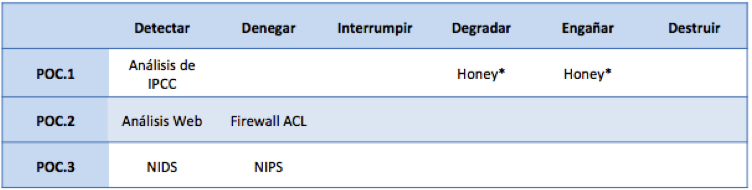}
\caption{CDA a nivel táctico (II)}
\end{figure} 

El nivel operacional alcanza en particular a los niveles medio de mando cuya responsabilidad radica en establecer los recursos tecnológicos y de seguridad necesarios como soporte de la misión y el logro de los objetivos. Cuanto más se sepa sobre los objetivos y las capacidades de un adversario, mejor se estará preparado para desarrollar un defensa proactiva. El nivel operacional del ciberconflicto se relaciona directamente con los eslabones: a) POC.1 Gestión de Campañas, b) POC.2 Militarización, y c) POC.3 Validación de objetivos; del MSOR. \newline

\textbf{Análisis y correlación de campañas}
 
El análisis de campañas permite detectar patrones comunes de comportamiento al analizar y correlacionar diferentes CDA. La Figura 9 evidencia el análisis de campaña en base a la correlación de dos CDA vinculados a OC.

\begin{figure}
\centering
\includegraphics[scale=1]{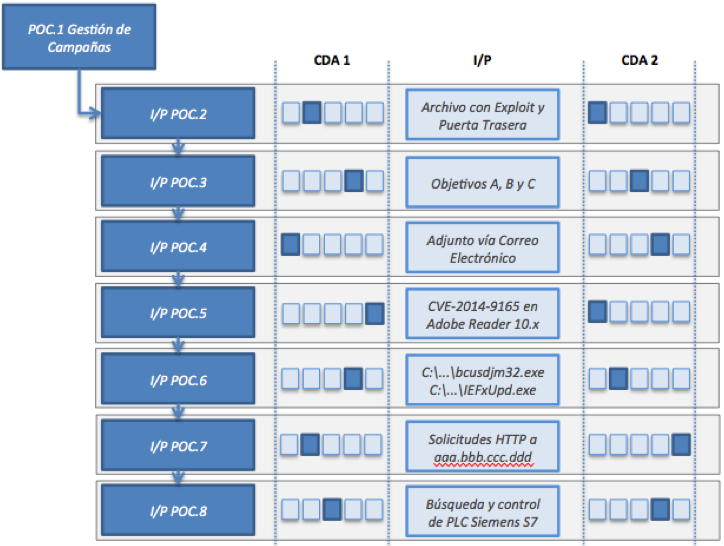}
\caption{Correlación de CDA}
\end{figure}

\emph{Indicadores y precursores clave de campaña:} Son aquellos que resultan más confiables, y sobre los cuales se sustenta la priorización y el desarrollo de los CDA. \newline

Los indicadores y precursores clave resultan menos volátiles que el resto y por consiguiente conforman la base de análisis. En este sentido, la calidad de la información derivada de este tipo de datos permite explotar la persistencia del adversario en beneficio propio. Dos recursos de valor aportados por el análisis de campañas: 

\begin{enumerate}
\item Sobre la atribución: El análisis de campañas puede aportar una herramienta para identificación positiva.
\item Sobre los objetivos del adversario: La determinación de los objetivos del adversario, de su misión, requiere de la determinación de tendencias en la selección de objetivos.
\end{enumerate}

La Figura 10 presenta un ejemplo de indicadores y precursores clave.

\begin{figure}
\centering
\includegraphics[width=\linewidth]{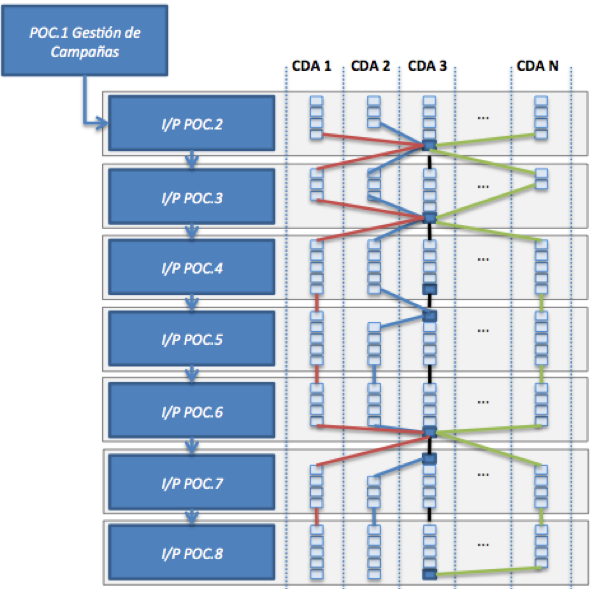}
\caption{Ejemplo de indicadores clave en una campaña}
\end{figure}

\subsection{Nivel estratégico del ciberconflicto }

\emph{Nivel estratégico:} Es el nivel en el que una organización, como tal o como parte de un grupo, establece los objetivos estratégicos de seguridad y la orientación a seguir, y hace uso de los recursos de los que dispone para conseguirlos. \newline

Este nivel conceptual del ciberconflicto involucra la toma de decisiones en los máximos niveles de responsabilidad de la organización y probablemente, para los casos en que se involucre infraestructura crítica, a organismos del estado tanto civiles como militares. Aún en tiempos de paz, la información aportada por los indicadores y precursores clave aportan datos fundamentales para visualizar los cambios en los escenarios de riesgo asociado a ciertas amenazas. \newline

El nivel estratégico alcanza a los siguientes componentes del MSOR:  a) PSC.1 Situación interna y geopolítica, b) PSC.2 Gestión de Riesgos, c) PSC.3 Desarrollo de Capacidades, d) PSC.4 Desarrollo Legal Nacional e Internacional, e) PSC.5 Diplomacia, f) PSC.6  Declaración del Conflicto, g) PPE.1 Inicio de la planificación, h) PPE.2 Análisis de la misión, i) PPE.3 Desarrollo de los CDA, j) PPE.4 Simulación y análisis de CDA, k) PPE.5 Comparación de CDA, l) PPE.6 Selección de CDA, y m) PPE.7 Desarrollo del plan.

El análisis del contexto de información asociado a los proceso PSC y PPE debe aportar la inteligencia necesaria para le modelo en base a interrogantes tales como:  

\begin{itemize}
\item ¿Quien es el adversario y a que obedece su comportamiento a largo plazo?
\item ¿Cuales son los blancos potenciales? 
\item ¿Qué tan importantes o valiosos son estos recursos?
\item ¿Qué tan bien se está protegiendo a estos recursos en la actualidad?
\item ¿Que debilidades tenemos que puedan ser explotadas por el adversario?
\item ¿En que jurisdicciones se desarrolla el conflicto?
\item ¿Cual es la autoridad en dichas jurisdicciones?
\item ¿Cual es el marco legal aplicable al conflicto?
\end{itemize}

El nivel estratégico requiere además que se consideren las cuestiones vinculadas a las infraestructuras críticas de la nación que con llevan una complejidad elevada desde la perspectiva de integración y cooperación debido a la naturaleza mixta dada por las divergencias entre propiedad y operación de las mismas.

\section{Justificación del modelo desde la lógica del C2}

En el presente capítulo se ofrecerá una justificación de la propuesta de CSI desde la perspectiva de comando y control (C2) en base a un análisis muy elemental soportado por los modelos OODA y de Brehmer para sistemas de control. El modelo OODA de Boyd \cite{b13} es utilizado como base para la generación de doctrinas y sistemas de C2 en el mundo. La Figura 11 muestra su estructura fundamental:

\begin{figure}
\centering
\includegraphics[width=\linewidth]{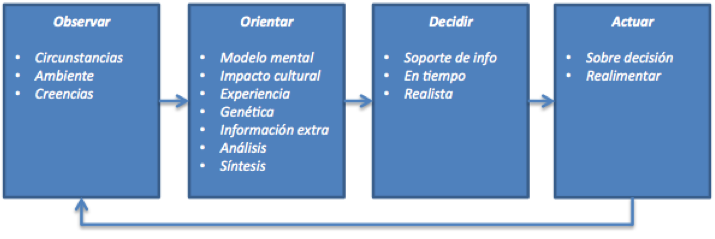}
\caption{Modelo OODA de Boyd}
\end{figure}  

OODA constituye un modelo conceptual sobre el cual es posible pensar el conflicto. Responde funcionalmente a la idea de un bucle realimentado, que si bien es criticado por algunos \cite{b14} \cite{b15}, a los fines del presente trabajo resulta suficiente por lo que se aplicará al ciberespacio para generar una abstracción de las generalidades de la teoría del C2. A la ventaja aportada por el ciberespacio en términos de observación y orientación por el concepto de GCR desde lo ofensivo, la CDI plantea una estrategia antagónica de ciberdefensa centrada en el desarrollo de interferencia de información en base a la utilización de la CCI para limitar o alterar la información de contexto que el adversario necesita para actuar. La interferencia, para que resulte útil en términos de afectación del sistema C2, debe ser lograda con la velocidad necesaria, hecho que se ajusta a la teoría general del conflicto en el que la velocidad es un factor clave de éxito según Liddel Hart \cite{b16}. La CCI puede pensarse como un instrumento destinado a penetrar el bucle OODA del adversario en las fases de observación y orientación con el fin de lograr ventajas, tanto en conflictos simétricos como asimétricos. \newline

El modelo OODA posee una limitación fundamental desde la perspectiva del C2 moderno dado por la falta de precisión en la definición de los retardos generados a nivel sistémico ya que solo considera el tiempo asociado a la toma de decisiones. Para salvar esta limitación y hacer aún más robusta la justificación del modelo de CSI planteado se recurre al modelo de control aportado por Brehmer  \cite{b17} en el que se consideran todos los retardos presentes en todo sistema de control, a saber: \textit{tiempo muerto, tiempos constantes, retardos de información y tiempo de decisión}.

\begin{figure}
\centering
\includegraphics[width=\linewidth]{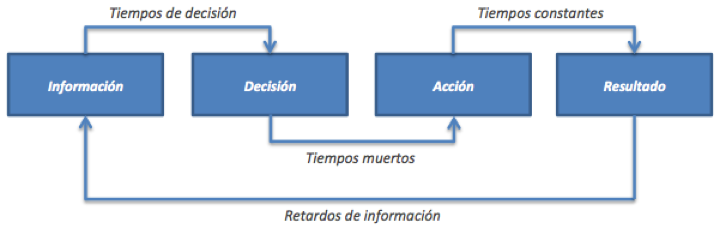}
\caption{Modelo de control de Brehmer}
\end{figure}

Según este modelo, la interferencia de información lograda por la CCI sobre la estructura de C2 del adversario puede tener efecto sobre otras fases de control, más allá de la de toma de decisión. Por ejemplo, podría producirse interferencia en forma de alteración de información sobre los datos asociados a la fase de resultados del modelo, retardando o engañando al adversario en relación a los resultados logrados por medio de las operaciones desarrolladas. \newline

\section{Conclusiones}

Como conclusión puede argumentarse que independientemente de la fase del MSOR con la que se esté tratando para lograr interferir en el bucle de decisión del adversario, y a diferencia de algunas concepciones tradicionales sobre el conflicto como la de Clausewitz  \cite{b18} en la que la velocidad solo debe darse en el ataque y no en la defensa, la defensa del ciberespacio requiere ser más rápido que el adversario. Este hecho se enfatiza en escenarios de conflicto asimétricos y contra amenazas de tipo APT. La propuesta de CSI busca afectar los tiempos en los bucles de control y así alterar de manera adversa las cuestiones de sincronización vinculadas a los esquemas de C2. \newline

El presente trabajo presenta un modelo de Ciberseguridad Inteligente (CDI) sustentado en el desarrollo de una Capacidad de Ciberinteligencia (CCI) como complemento a las capacidades tradicionales asociadas a la seguridad operativa y a la gestión de incidentes de seguridad en el ciberespacio. A tal fin se presenta el Modelo de Sistema Ofensivo de Referencia (MSOR) el cual es utilizado para pensar el ciberconflicto a todo nivel, desde una perspectiva coordinada y sincronizada con el resto de las fuerzas tradicionales en el marco del conjunto. La CSI se logra por medio de la formalización teórica y el desarrollo de una CCI destinada a detectar, denegar, interrumpir, degradar, engañar y destruir las capacidades de inteligencia del adversario, por medio de la aplicación de una estrategia centrada en la interferencia de información vinculada a los sistemas C2, penetrando sus bucles de decisión con la velocidad necesaria, a fin de lograr que este se posicione en una postura reactiva. La propuesta posee otra ventaja significativa y es que favorece el desarrollo de campañas por medio de la aplicación de un marco ágil en el contexto de asimetría de conflicto, aspecto a considerar al momento de tener que enfrentar a adversarios que representen amenazas de tipo APT. Finalmente, a diferencia de los modelos de defensa planteados clásicamente, el concepto de CSI sugiere que la ventaja en el conflicto puede ser obtenida por el defensor y no siempre por el atacante.



\end{document}